# What is- and what is not- Electromagnetically-Induced-Transparency in Whispering-Gallery-Microcavities

Bo Peng[1], Şahin Kaya Özdemir[1*], Weijian Chen[1], Franco Nori[2], and Lan Yang[1*]

**Affiliations:**

[1]Department of Electrical and Systems Engineering, Washington University, St. Louis, MO 63130, USA

[2]CEMS, RIKEN, Saitama 351-0198, Japan

*Correspondence to:  yang@ese.wustl.edu, ozdemir@ese.wustl.edu

**Abstract**

Electromagnetically-induced-transparency (EIT) and Autler-Townes splitting (ATS) are two prominent examples of coherent interactions between optical fields and multilevel atoms. They have been observed in various physical systems involving atoms, molecules, meta-structures and plasmons. In recent years, there has been an increasing interest in the implementations of all-optical analogues of EIT and ATS via the interacting resonant modes of one or more optical microcavities. Despite the differences in their underlying physics, both EIT and ATS are quantified by the appearance of a transparency window in the absorption or transmission spectrum, which often leads to a confusion about its origin. While in EIT the transparency window is a result of Fano interference among different transition pathways, in ATS it is the result of strong field-driven interactions leading to the splitting of energy levels. Being able to tell objectively whether a transparency window observed in the spectrum is due to EIT or ATS is crucial for clarifying the physics involved and for practical applications. Here we report a systematic study of the pathways leading to EIT, Fano, and ATS, in systems of two coupled whispering-gallery-mode (WGM) microtoroidal resonators. Moreover, we report for the first time the application of the Akaike Information Criterion discerning between all-optical analogues of EIT and ATS, and clarifying the transition between them.

**Introduction**

Coherent interactions of light fields with multilevel atomic systems can dramatically modify the optical response of the medium, via quantum interferences between various excitation pathways, or via the splitting of energy levels into dressed states by strong coupling fields. The process known as electromagnetically-induced-transparency (EIT)[1,2,3,4] is the result of interference effects, namely the Fano interferences[5,6] which require coupling of a discrete transition to a continuum. EIT eliminates a resonant absorption by creating a narrow transparency window with subnatural linewidth in an otherwise opaque material, while leading to enhanced nonlinearity associated with near-resonant effects. These features of EIT have led to a rich variety of applications including (but not limited to) ultra-slow light propagation[7], storage of light pulses[8,9], transmission of weak light without dissipation through otherwise optically-dense media, and nonlinear optics with weak light. Autler-Townes splitting (ATS)[10], on the other hand, is a process involving strong field-induced splitting of energy levels and is not associated with interference effects; yet it creates a transparency window due to the doublet structure in the absorption profile. It has been used for alignment of angular momentum, measurements of transition dipole moments, and quantum control of spin–orbit perturbations.

Coherent processes leading to EIT and ATS have been observed in: atomic gases[7,11], a variety of atomic and molecular systems[12], solid-state systems[13], superconducting circuits[14,15], plasmonics[16], metamaterials[17], optomechanical systems[18,19], inductively or capacitively-coupled resonant electronic circuits[20], photonic crystals[21], and whispering-gallery-mode (WGM) optical resonators[22,23,24,25,26,27,28,29]. A brief illustration of different systems in which EIT and ATS have been studied are given in **Fig.1** and **Table 1,** showing the analogies among them. The existence of these processes in plasmonic, metamaterial, photonic crystal, and WGM microresonators is critical for on-chip coherent control and manipulation of light at room temperatures. EIT and ATS in these systems do not suffer from experimental complexities and difficulties that are common in implementations in solid-state and atomic media (e.g., a low temperature environment, the need for stable and well-controlled lasers to match the atomic transitions, or propagation-scaling limitations as a result of control-field absorption).

Among the many physical systems, WGM microresonators have been the most fruitful platforms to study various theoretical and experimental aspects of classical and all-optical analogues of EIT and ATS for on-chip implementations. Fano resonances and EIT have been observed in: a single silica microsphere side-coupled to a fiber taper[26], a polydimethyl-siloxane- (PDMS-) coated silica microtoroid with two simultaneously excited WGMs[27], two directly-coupled silica microspheres[23,28], two silicon microrings indirectly coupled via two waveguides in an add-drop filter configuration[24], and a system of a microdisk and a microtoroid that are indirectly coupled to each other through a fiber taper[29]. In all of these EIT implementations, two WGMs of high and low quality factors (Q) are coupled, having zero-detuning in their resonance frequencies, and destructive interference of the optical pathways cancels the absorption leading to a narrow peak in the transmission spectra. If a frequency detuning is introduced, the transmission spectra show sharp asymmetric Fano resonances. ATS has been observed in: directly-coupled silica microspheres, directly-coupled silica microtoroids[30], and hybrid systems formed by directly coupling silica microtoroids and microspheres with PDMS-coated silica microtoroids[31]. In these systems of coupled-WGM-resonators, ATS originates from the lifting of the frequency-degeneracy of the eigenmodes, and hence their splitting into two resonance modes due to the strong coupling between the resonators. The spectral region between the split modes corresponds to a transparency window. In the context of high-performance sensing, a scatterer-induced coupling (e.g., single nanoparticle, single virus, etc.) between the frequency-degenerate clockwise and counter-clockwise rotating modes of a WGM resonator has been shown to lift the degeneracy leading to mode-splitting and ATS[32,33]. This has been used to detect and measure the polarizability of nano-scale objects with single particle resolution, as well as to directly measure the Purcell factor[34].

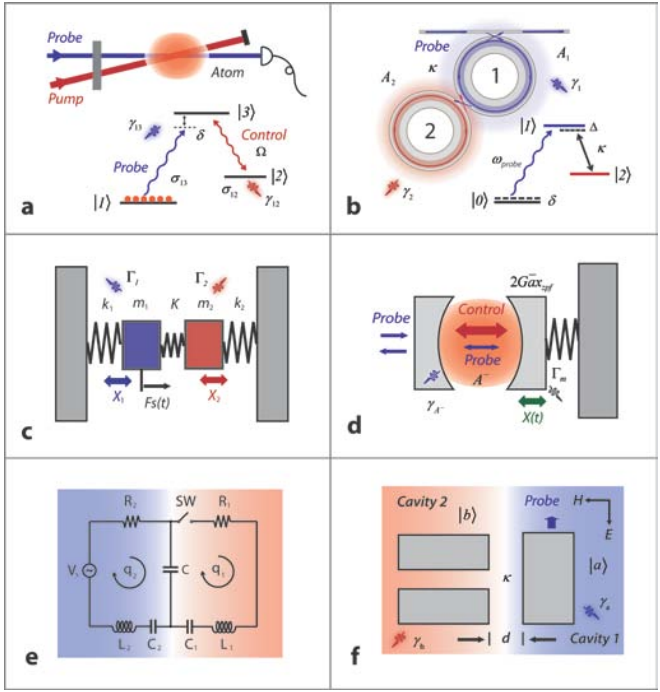

*Fig.1. Different platforms used for realizing electromagnetically induced transparency.* *(a) Atomic ensembles with $\Lambda$-type three-level atoms: Ground state $|1\rangle$; dark state $|2\rangle$; excited state $|3\rangle$; density matrix elements $\sigma_{12}$ and $\sigma_{13}$; pump field Rabi frequency $\Omega$; dephasing rates $\gamma_{12}$ and $\gamma_{13}$; detuning $\delta$. (Ref.15). (b) Coupled system of microresonators: Intracavity field amplitudes $A_1$ and $A_2$; cavity coupling strength $\kappa$; decay rates $\gamma_1$ and $\gamma_2$ of the optical modes $A_1$ and $A_2$; resonance detuning $\Delta$; detuning of the probe field $\delta$. (Ref.22). (c) Mechanical systems: spring constants $K, k_1, k_2$; masses $m_1$ and $m_2$; displacements $X_1$ and $X_2$ of mass 1 and mass 2; mechanical dampings $\Gamma_1$ and $\Gamma_2$; driving force $F_s(t)$. (Ref.20). (d) Optomechanical systems: Intracavity field amplitude $A^-$; mechanical displacement $X$; optomechanical coupling strength $2G\bar{a}x_{zpf}$; decay $\gamma_{A^-}$ of $A^-$; mechanical damping $\Gamma_m$. (Ref.18). (e) Electronic circuits: Inductors $L_1$ and $L_2$; capacitors $C, C_1$ and $C_2$; resistors $R_1$ and $R_2$; voltage source $V_s$; charges $q_1$ and $q_2$ in the resonance circuits; switch SW. (Ref.20). (f) Plasmonics: The radiative plasmonic state $|a\rangle$; the dark plasmonic state $|b\rangle$; plasmonic coupling strength $\kappa$; damping rates $\gamma_a$ and $\gamma_b$. (Ref.16). Correspondences and analogies among these system-specific parameters are shown in **Table 1**.*

*Table 1* *Correspondences among parameters of various systems (**Fig. 1**) in which EIT and ATS have been experimentally observed.*

|  | Atomic | Optomechanics | Resonators | Electronics | Mechanical | Plasmonics |
|---|---|---|---|---|---|---|
| Density matrix element (Radiative state) | $\sigma_{13}$ | $A^-$ | $A_1$ | $q_1$ | $X_1$ | $|a\rangle$ |
| Density matrix element (Dark state) | $\sigma_{12}$ | $X$ | $A_2$ | $q_2$ | $X_2$ | $|b\rangle$ |
| Coupling strength (Rabi Frequency) | $\mu_{23}\xi_c/\hbar$ | $2G\bar{a}x_{zpf}$ | $\kappa$ | $\sqrt{1/(L_2 C)}$ | $\sqrt{K/m_1}$ | $\kappa$ |
| Dephasing rate | $\gamma_{13}$ | $\gamma_{A^-}$ | $\gamma_1$ | $R_1/L_1$ | $\Gamma_1$ | $\gamma_a$ |
| Dephasing rate | $\gamma_{12}$ | $\Gamma_m$ | $\gamma_2$ | $R_2/L_2$ | $\Gamma_2$ | $\gamma_b$ |

The effects of ATS in the absorption profile of the material system resemble that of EIT, in that both of the processes display a transparency window, i.e., a reduction in the absorption profile in the presence of a coupling field. This resemblance also appears in their all-optical-classical-analogues in coupled optical resonators. This similarity has led to much confusion[14,35,36] and many discussions[15,37,38,39,39,40] on how to discriminate between EIT and ATS processes just by looking at the absorption/transmission spectra obtained in an experiment, without prior knowledge on the system. The sharpness of the dip in the absorption and the imaginary part of the susceptibility (or the peak in the transmission) has long been used as an intuitive and informal criterion to judge whether EIT takes place or not. However, such a test is very subjective and has generated confusion. For example, a peak in the transmission spectrum of a system is in general much sharper than the dip in the absorption spectrum of the same system. Therefore, one has to first decide whether to focus on the sharpness of the transmission or the sharpness of susceptibility/absorption. Moreover, system-specific parameters affect the output spectra. Namely, relaxation rates, coupling strengths, cleanliness of the samples etc. differ significantly among different systems and determine the sharpness in the measured output spectrum. For example, these parameters in superconducting systems are at least one to two orders-of-magnitude less than their atomic counterparts, and hence the dip in the susceptibility of superconducting systems is less sharp than that of alkaline atoms[41]. Therefore, objective methods and tests are needed to make claims of EIT in different physical systems. Since identifying whether an experiment involves EIT or ATS is important for applications, objective methods to discriminate between them have been sought.

Many experimental and theoretical studies with different configurations of three-level atomic systems have been carried out to identify the required conditions to observe EIT or ATS. Recently Anisimov, Dowling and Sanders[15] have proposed to use the Akaike Information Criterion (AIC)[42] as an objective test to discern EIT from ATS in experimentally obtained absorption or transmission spectra, and to identify the spectra from which one cannot derive a conclusive result on whether EIT or ATS has played a role. They have successfully applied this test to an experiment with a 1D superconducting transmission line coupled to a flux qubit, concluding that the reported data[14] do not support the claim of EIT. In a recent study[43], Giner *et al.* used AIC on data obtained in their experiment involving an ensemble of cold cesium atoms and demonstrating the suitability of the criterion to evaluate experimental data objectively. Oishi *et al.*[44] investigated the transient response of coupled optical fiber ring resonators to a square-pulse input, attempting to find precursors of EIT and ATS. This study revealed that, when the system was prepared for EIT, a sharp spike was seen in the transient response; whereas when the system was prepared for ATS, an oscillatory signal attributed to coherent energy exchange between the resonators was observed. Thus they suggest that the transient response of a system can be used to discern EIT from ATS in any physical system under consideration. It is to be noted here that AIC discriminates EIT from ATS in the frequency domain, whereas the transient response method discriminates them in the time domain.

Up to this date, there was no study being carried out to discern EIT from ATS in coupled WGM optical resonators using the AIC. Here, we systematically investigate Fano resonances, EIT and ATS in a system of two coupled WGM microtoroid resonators, identify the transition from EIT to ATS (and vice versa), and use AIC to discern EIT from ATS from the experimentally-obtained transmission spectra. Our results show the suitability of the AIC for discriminating EIT from ATS in systems of coupled optical resonators.

**Results**

**Experimental set-up.** Our system consists of two directly-coupled silica microtoroidal WGM resonators μR1 and μR2, with μR1 coupled to a fiber-taper. (**Fig. 2**). We fabricated the silica microtoroids at the edges of two separate silicon wafers, such that when the wafers were brought closer to each other, the microtoroids begin exchanging energy. The wafers were placed on separate nanopositioning systems so that the distance between the microtoroids was finely tuned to control the coupling strength $\kappa$ between them. The coupling strength $\kappa$ decreases exponentially with increasing distance. The probe light in the 1550 nm band from a narrow linewidth tunable laser was coupled into a WGM of μR1 via the fiber taper. The same fiber taper was also used to couple out the light from the WGM. The output light was then sent to a photodetector connected to an oscilloscope, to obtain the transmission spectra as the wavelength of the input light was linearly scanned. Fiber-based polarization controllers were used to set the polarization of the input light for maximal coupling into WGMs. A thermo-electric-cooler was placed under one of the wafers so that resonance frequency of the WGM of interest in a microtoroid could be tuned via the thermo-optic effect, to control the frequency detuning of the chosen WGMs in the two microtoroids. A tuning range of 8 GHz was achieved. The microtoroids supported many WGMs in the same band but with different quality factors Q, which is the signature of the amount of loss or dissipation (the lower the loss, the higher the Q and the narrower the linewidth of the resonance mode). This allowed us to investigate the effects of Q of the selected modes on the Fano, EIT and ATS processes by choosing WGM-pairs with different Q-contrasts. In addition to the ability of choosing different WGM-pairs, our set-up allowed us to investigate Fano, EIT and ATS processes and the transitions among them by steering the system independently via the coupling strength or the frequency detuning between the selected WGMs. In our experiment, we selected three different sets of WGM pairs with the intrinsic quality factors $(Q_{\mu R1}, Q_{\mu R2})$ of $(1.91 \times 10^5, 7.26 \times 10^7)$, $(1.78 \times 10^6, 4.67 \times 10^6)$ and $(1.63 \times 10^6, 1.54 \times 10^6)$. Note that the intrinsic Q includes all the losses (e.g., material, radiation, scattering) except the coupling losses. Since the probe light is input at the µR1 side with a fiber taper, the Q of the µR1 is smaller than the above intrinsic Q values due to the additional coupling losses (i.e., µR1 has more loss than µR2).

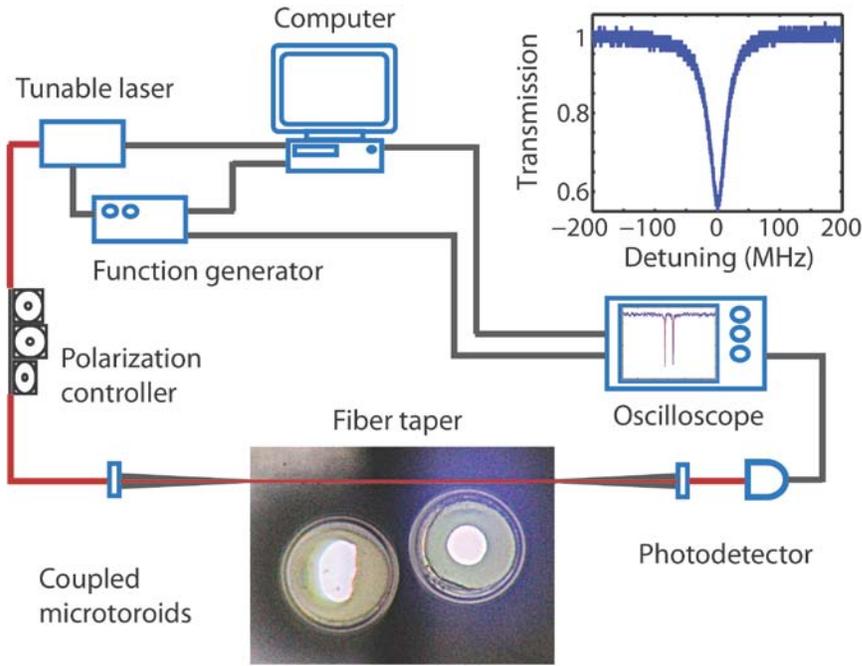

*Fig.2. Experimental configuration for the coupled whispering-gallery-mode (WGM) microcavities system.* *Setup with a microscope image of the coupled microtoroid system. The system consists of two directly coupled WGM resonators (μR1 and μR2), and a fiber-taper waveguide coupled to μR1. The probe laser is in the 1550 nm band. Inset: Typical transmission spectra showing the resonant mode in μR1.*

**Analogy between coupled optical resonators and three-level atoms.** Here we will elucidate the analogy between atomic and photonic coherence effects leading to EIT and ATS. Using coupled-mode-theory, we find the equations of motion for the complex intracavity field amplitudes $A_1$ and $A_2$ in the steady-state as

$$(\delta_1 + i\gamma_1 / 2) A_1 - \kappa A_2 = -i\sqrt{\gamma_c} A_p \tag{1}$$

$$(\delta_2 - i\gamma_2 / 2) A_2 - \kappa A_1 = 0 \tag{2}$$

where $\gamma_1 = \gamma_1' + \gamma_c$ and $\gamma_2$ denote the total losses in µR1 and µR2, respectively, $\gamma_1'$ is the intrinsic loss of µR1 and $\gamma_c$ is the coupling loss between the fiber-taper and µR1, $\delta_1 = \omega - \omega_1$ and $\delta_2 = \omega - \omega_2$ denote the detuning between the frequency $\omega$ of the probe light field $A_p$ and the resonance frequencies $\omega_1$ and $\omega_2$ of the WGMs, and $\kappa$ is the coupling strength between the WGMs. In the EIT and ATS experiments we set $\omega_1 = \omega_2 = \omega_0$ via the thermo-optic effect by thermally tuning the frequency of one of the WGMs to be equal to the frequency of the other. Consequently, for the degenerate frequencies $\omega_1$ and $\omega_2$ we have $\Delta = \omega_2 - \omega_1 = 0$, and in the rotated frame ($\omega_0 \to 0$) we have $\delta_1 = \delta_2 = \omega$. Also note that in the system depicted in Fig. 1c and Fig.2, the input and output ports are at the side of µR1, hence the output field is given as $A_{out} = A_p + \sqrt{\gamma_c} A_1$ where the intracavity field $A_1$ can be written as $A_1 = i\sqrt{\gamma_c} A_p \chi$ with

$$\chi = \frac{(\omega + i\alpha_2)}{\kappa^2 - (\omega + i\alpha_1)(\omega + i\alpha_2)} \tag{3}$$

where we used $\alpha_k = i\omega_k + \gamma_k/2$, with $k = 1, 2$. This solution $\chi$ has a form similar to the response of an EIT medium (three-level atom) to a probe field. Then, we can write the normalized transmission $T = |A_{out}/A_p|^2$ as

$$T = 1 + \gamma_c^2 |\chi|^2 - 2\gamma_c \chi_i \tag{4}$$

where $\chi_i$ is the imaginary part of $\chi$. Since $1 \gg \gamma_c^2 |\chi|^2$ and $\gamma_c^2 |\chi|^2 \ll \gamma_c \chi_i$ we can re-write the transmission as

$$T = 1 - 2\gamma_c \chi_i \tag{5}$$

Thus it is sufficient to analyze the behavior of $\chi_i$ to understand the conditions leading to EIT or ATS. This is similar to considering the imaginary part of the susceptibility which determines the absorption of a probe in an atomic system.

The eigenfrequencies of this coupled system can be found from the denominator of Eq. (3) and are given as $\omega_\pm = (-i\alpha_1 - i\alpha_2 \pm \beta)/2$, with $\beta^2 = 4\kappa^2 - (\alpha_1 - \alpha_2)^2$. This reveals a transition at the threshold coupling strength $2|\kappa_T| = |\alpha_1 - \alpha_2| = \gamma_1 - \gamma_2$, where we have used the fact that in our system $\gamma_1 > \gamma_2$, as stated in the previous section. We define the regimes where $\kappa < \kappa_T$ and $\kappa > \kappa_T$ as the weak- and strong-driving regimes, respectively, and $\kappa = \kappa_T$ as the transition point. Using the eigenfrequencies, we can re-write the expression in Eq.(3) as

$$\chi = \frac{\chi_+}{\omega - \omega_+} + \frac{\chi_-}{\omega - \omega_-} \tag{6}$$

where $\chi_\pm = \mp(\omega_\pm + i\alpha_2)/\beta = -1/2 \pm i\xi/\beta$ satisfying $\chi_+ + \chi_- = -1$ and $\xi = (\gamma_1 - \gamma_2)/4$.

*(a) Weak-driving regime ($\kappa < \kappa_T$).* In this regime, $\beta$ is imaginary, that is $\beta = i\beta_i$ and $\text{Re}(\beta) = \beta_r = 0$. This leads to a real $\chi_\pm$ (i.e, $\text{Im}(\chi_\pm) = \chi_{\pm i} = 0$), with $\text{Re}(\chi_\pm) = \chi_{\pm r} = -1/2 \pm \xi/|\beta|$, and imaginary eigenfrequencies with $\text{Re}(\omega_\pm) = \omega_{\pm r} = 0$ and $\text{Im}(\omega_\pm) = \omega_{\pm i} = -\zeta \pm |\beta|$, where $\zeta = (\gamma_1 + \gamma_2)/4$. Thus the supermodes have the same resonance frequencies and are located at the center of the frequency axis, but have different linewidths quantified by their imaginary parts. The imaginary parts of $\chi$ are then given by

$$\chi_i = \frac{\omega_{+i}\chi_{+r}}{\omega^2 + \omega_{+i}^2} + \frac{\omega_{-i}\chi_{-r}}{\omega^2 + \omega_{-i}^2} \tag{7}$$

which consists of two Lorentzians centered at $\omega = 0$ with different signs (i.e, the first term in Eq. is negative whereas the second term is positive). The transmission in this regime becomes

$$T_{EIT} = 1 - 2\gamma_c \chi_i = 1 - 2\gamma_c \left[ \frac{\omega_{+i}\chi_{+r}}{\omega^2 + \omega_{+i}^2} + \frac{\omega_{-i}\chi_{-r}}{\omega^2 + \omega_{-i}^2} \right] = 1 - \frac{C_1}{\omega^2 + \Gamma_1^2} + \frac{C_2}{\omega^2 + \Gamma_2^2} \tag{8}$$

where all the parameters whose values cannot be determined precisely are incorporated into the coefficients $C_k$ and $\Gamma_k$ that can be used as free parameters to perform curve-fitting to experimentally-obtained transmission spectra. Clearly, the opposite signs of the Lorentzians lead to a destructive interference which results in a transmission profile exhibiting a transparency window similar to that of EIT.

*(b) Strong-driving regime ($\kappa \gg \kappa_T$)*. In this regime $\beta = 2\kappa$ is real (i.e., $\beta_i = 0$ and $\beta_r = 2\kappa$) implying $\omega_\pm = -i\zeta \pm \kappa$, that is the resonances are located at frequencies $\pm\kappa$ with a spectral distance of $2\kappa$ between them. The resonance linewidths are quantified by $\text{Im}(\omega\pm) = -\zeta$. Approximating $\chi_\pm$ as $\chi_\pm = -1/2$ we find the imaginary part of $\chi$ as

$$\chi_i = \frac{1}{2}\left[\frac{\zeta}{(\omega-\kappa)^2 + \zeta^2} + \frac{\zeta}{(\omega+\kappa)^2 + \zeta^2}\right] \tag{9}$$

which implies that $\chi_i$ is the sum of two same-sign Lorentzians centered at $\pm\kappa$. The transmission in this regime is then given by

$$T_{ATS} = 1 - \gamma_c \left[\frac{\zeta}{(\omega-\kappa)^2 + \zeta^2} + \frac{\zeta}{(\omega+\kappa)^2 + \zeta^2}\right] = 1 - \frac{C}{(\omega-\delta)^2 + \Gamma^2} - \frac{C}{(\omega-\delta)^2 + \Gamma^2} \tag{10}$$

where $C$, $\Gamma$ and $\delta$ are the free parameters used in curve-fitting to experimentally obtained transmission spectra. Clearly, the transmission in this strong-driving regime presents a symmetric doublet spectra and the observed transparency is due to the contribution of two Lorentzians.

*(c) Intermediate-driving regime ( $\kappa > \kappa_T$ ).* In this regime $\beta = \beta_r$ is real (i.e., $\beta_i = 0$). This leads to complex eigenfrequencies $\omega_\pm = \left(-i\gamma_1 - i\gamma_2 \pm 2\beta_r\right)/4$ and complex $\chi_\pm = -1/2 \pm i(\gamma_1 - \gamma_2)/4\beta_r$. Thus the supermodes have different resonance frequencies located at $\pm \beta_r/2$, but have the same linewidths quantified by their imaginary parts $\text{Im}(\omega_\pm) = \omega_{\pm i} = (-\gamma_1 - \gamma_2)/4$. Consequently, we have

$$\chi_i = \frac{(\omega - \omega_{+r})\chi_{+i} + \omega_{+i}\chi_{+r}}{(\omega - \omega_{+r})^2 + \omega_{+i}^2} + \frac{(\omega - \omega_{-r})\chi_{-i} + \omega_{-i}\chi_{-r}}{(\omega - \omega_{-r})^2 + \omega_{-i}^2} \tag{11}$$

and

$$T_{EIT/ATS} = 1 - \left[\frac{(\omega - \varepsilon)C_1}{(\omega - \varepsilon)^2 + \Gamma^2} - \frac{(\omega + \varepsilon)C_1}{(\omega + \varepsilon)^2 + \Gamma^2}\right] - \left[\frac{C_2}{(\omega - \varepsilon)^2 + \Gamma^2} + \frac{C_2}{(\omega + \varepsilon)^2 + \Gamma^2}\right] \tag{12}$$

The expression in the second bracket of Eq. (12) is the sum of two Lorentzians, similar to the expression obtained for the strong-driving regime in Eq. (10), implying the contribution of ATS. The expression in the first bracket corresponds to the interference term, and can be controlled by choosing the loss of the coupled modes. For example, choosing two modes satisfying $\gamma_1 = \gamma_2$ will lead to $C_1 = 0$, and hence the expression $T_{EIT/ATS}$ will become the same as $T_{ATS}$. This implies that to observe ATS, the linewidths (i.e., Q) of the coupled WGMs should be very close to each other as will be demonstrated in the experiments discussed below.

The theoretical shapes and more detailed discussions of these operating regimes are given in the Appendix.

**Experimental demonstration of Fano resonances and EIT in coupled WGM microresonators.** EIT is a result of strong Fano interferences, and takes place when a high-Q WGM of one microresonator is directly-coupled to a low-Q WGM of a second microresonator with zero-detuning in their resonance frequencies. In order to demonstrate this, we chose a low-Q mode in μR1 ($Q_{\mu R1} = 1.91 \times 10^5$) and a high-Q mode in μR2 ($Q_{\mu R2} = 7.26 \times 10^7$). We then set the distance between the resonators such that the coupling strength was smaller than the total loss of the system. At this point, we continuously tuned the frequency of the high-Q mode in µR2 such that it approached to the frequency of the low-Q mode in µR1. As the frequency-detuning between the modes gradually decreased, the modes became spectrally overlapped. Consequently, we first observed an asymmetric Fano lineshape with the peak located closer to the lower-detuning side (**Fig. 3a**, upper panels), and then a transparency window appeared at zero-detuning $\omega_2 - \omega_1 = 0$ (**Fig.3a**, middle panel). The linewidth of the transparency windows was 5 MHz. The asymmetry of Fano resonances decreased as we approached to zero-detuning. As the frequency of the high-Q mode was further increased, detuning started to increase again leading to the emergence of Fano lineshapes whose peaks were also located closer to zero-detuning (**Fig. 3a**, lower panels). Finally, when the frequency was increased such that there was no overlap between the modes, Fano lineshapes were lost and we observed two independent Lorentzian lineshapes corresponding to the two modes in µR1 and µR2.

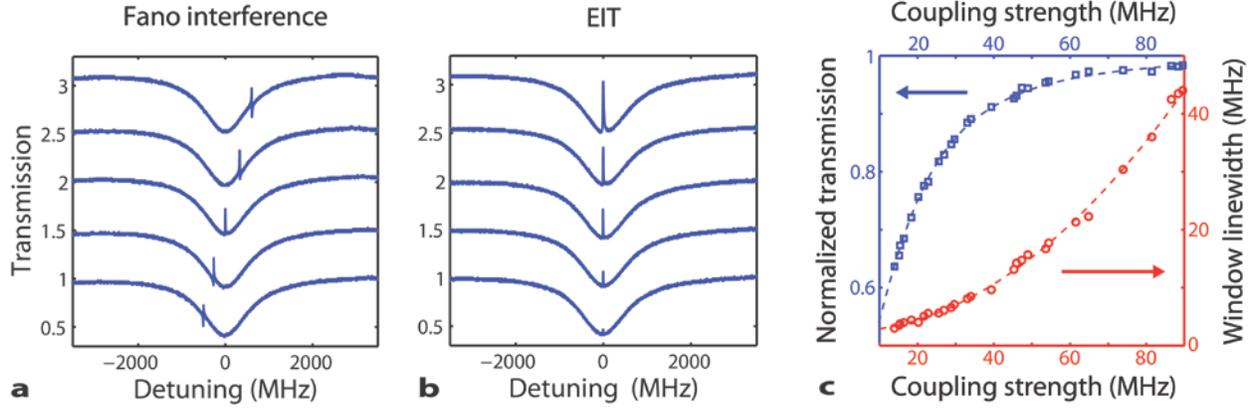

*Fig.3. Fano interference and Electromagnetically induced transparency (EIT) in coupled WGM microcavities. (**a**) Spectral tuning of the system from asymmetric Fano resonance to EIT. When the resonance modes of the microresonators have non-zero frequency-detuning, the transmission exhibits asymmetric Fano resonances. At zero-detuning, a transparency window appears. (**b**) Effect of coupling strength on the EIT spectra (i.e., zero detuning between resonance modes of the resonators). The coupling strength (increasing from the bottom to the top curve) depends on the distance between the resonators. (**c**) Effect of the coupling strength on the linewidth (red circles) and the peak transmission (blue squares) of the transparency window. The curves are the best fit to the experimental data.*

Next we studied the effect of coupling on the transparency window by first setting the frequency detuning of the low- and high-Q modes to zero, and then tuning the distance between the microtoroids (**Fig. 3b**). Note that the coupling strength here corresponds to the strength of the control field in atomic systems. We observed that as the coupling strength was increased (i.e., the distance between the microtoroids decreased) the transmission at the transparency window increased from 0.63 to 0.98 (**Fig. 3c**). During this process, the linewidth of the transparency window increased from 3 MHz to 43 MHz (**Fig. 3c**).

**Experimental demonstration of ATS in coupled WGM microresonators.** Contrary to EIT, ATS is not a result of Fano interferences but requires a strong coupling between two WGMs of similar Q. In order to demonstrate this, we chose the mode with $Q_{\mu R1} = 1.63 \times 10^6$ in µR1 and the mode $Q_{\mu R2} = 1.54 \times 10^6$ in µR2. We first tuned the resonance frequencies of the modes to have $\omega_1 = \omega_2$ (i.e., zero-detuning) when the microtoroids were sufficiently away from each other so that they could not exchange energies (i.e., no coupling). At this stage the transmission showed single resonance with Lorentzian lineshape (**Fig. 4a**, lowest panel). As we started to bring the resonators closer to each other (i.e, increased coupling strength), the single resonance split into two resonances whose spectral distance (i.e., mode splitting) increased with increasing coupling strength. For large coupling strengths, the transmission spectra were well-fitted by two Lorentzian resonances.

Next we chose two detuned resonance lines and set the coupling to strong-coupling regime (i.e., resonators are very close to each other). We observed that the split modes in the transmission were not symmetric (**Fig. 4b**, upper panel), and they had different transmission dips. This can be attributed to the unequal distribution of the supermodes in the two resonators. As we tuned the spectral distance between the WGMs by increasing the frequency of the mode in µR2, the split modes started to approach each other (i.e, decreasing mode-splitting) and the difference between their transmission dips decreased. At zero-detuning the resonances became symmetric, that is they are Lorentzian with the same linewidths and transmission dips (**Fig. 4b**, middle panel).

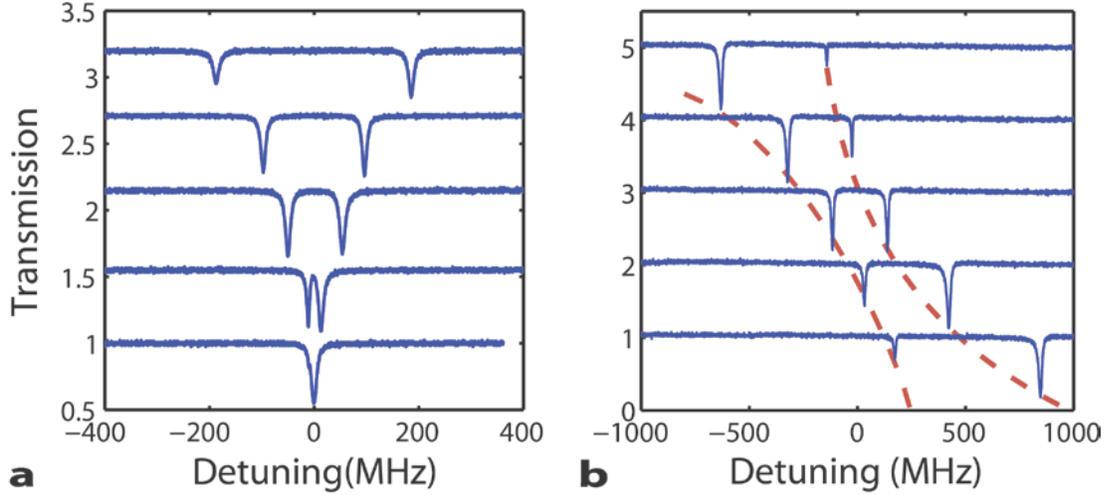

*Fig.4. Autler-Townes Splitting (ATS) and avoided-crossing in coupled WGM microcavities. (a) Effect of the coupling strength on the ATS spectra. During the process, the resonance modes have zero frequency-detuning, and as the coupling strength was increased (from the bottom to the top curve) the splitting increased. (b) Effect of the frequency-detuning of the coupled resonant modes on the ATS spectra. When the two resonant modes have non-zero detuning, the transmission exhibits asymmetric ATS (unequal transmission dips). As the frequency of one of the resonant modes is increased (frequency-detuning approaches to zero), the split modes start to approach each other and the transmission exhibits more symmetric ATS spectrum. With further increase of the frequency, the split modes exhibit avoided crossing. The smallest splitting was ~300 MHz, at which the frequency-detuning between the resonant modes was zero. Further increase of the resonance frequency then leads back to an asymmetric ATS.*

Here, the supermodes are equally distributed between the resonators. When the frequency of the mode in μR2 was further increased beyond the zero-detuning point, the modes repelled each other leading to an avoided-crossing during which they interchanged their linewidths and transmission dips (**Fig. 4b**, lower panels).

**Objectively discerning between EIT and ATS in coupled microresonators using Akaike Information Criteron (AIC).** As we have discussed in the previous sections, in the theoretical model and the experimental observations, EIT and ATS both display a transparency window in the transmission spectrum of coupled microresonators (i.e., similar to a three-level quantum system). However, EIT is due to Fano interference and hence requires coupling between a high-Q mode (i.e., discrete system) and a low-Q mode (i.e., continuum) in the weak-driving regime, while ATS is due to strong-coupling induced splitting of resonance modes and requires the interaction between modes with similar Q in the strong-driving regime. Thus depending on the relative Q's of the interacting modes and their coupling strength, the system of coupled microresonators can operate in either the EIT or the ATS regime. Discerning whether a transparency window in the transmission spectrum of the system of coupled resonators is the signature of EIT or ATS without apriori information on the Qs and coupling strength between the modes is crucial.

Here, we performed experiments in various conditions of our coupled-resonators system, obtained transmission spectra, and used the Akaike Information Criterion (AIC) proposed to discriminate between EIT and ATS in atomic systems to discern EIT or ATS. The AIC provides a method to select the best model from a set of models based on the Kullback-Leibler (K-L) distance between the model and the truth. The K-L distance quantifies the amount of information lost when approximating the truth. Thus, a good model is the one which minimizes the information loss and hence the K-L distance. Then AIC quantifies the amount of information lost when the model $\lambda_i$ with $k_i$ unknown parameters out of $N$ models is used to fit the data $x = x_1, x_2, \ldots, x_n$ obtained in the measurements, and is given as $I_i = -2\log L_i + 2k_i$, where $L_i$ is the maximum likelihood for the candidate model $\lambda_i$ and $2k_i$ accounts for the penalty for the number of parameters used in the fitting. Then the relative likelihood of the model $\lambda_i$ is given by the Akaike weight $w_i = e^{-I_i/2} / \sum_{j=1}^{N} e^{-I_j/2}$. In the case of least-squares and the presence of technical noise in the experiments, a fitness test using per-point (mean) AIC weight $\bar{w}_i = e^{-\bar{I}_i/2} / \sum_{j=1}^{N} e^{-\bar{I}_j/2}$, where $\bar{I}_i = I_i/n$. In our study involving only two models ($N = 2$), we can re-write $w_i$ and $\bar{w}_i$ as

$$w_{EIT} = \frac{e^{-I_{EIT}/2}}{e^{-I_{EIT}/2} + e^{-I_{ATS}/2}}, \qquad \bar{w}_{EIT} = \frac{e^{-I_{EIT}/2n}}{e^{-I_{EIT}/2n} + e^{-I_{ATS}/2n}} \qquad (13)$$

with $w_{EIT} + w_{ATS} = 1$ and $\bar{w}_{EIT} + \bar{w}_{ATS} = 1$.

In our experiments, we acquired ten-thousand data points ($n = 10,000$) to form a transmission spectrum at each setting of coupling-strength and frequency-detuning, and we repeated the measurements to obtain ten transmission spectra at each setting, to take into account the fluctuations and uncertainty in the measurements. We used $T_{EIT}$ and $T_{ATS}$ given in Eqs. 8 and 12, respectively, to fit the transmission spectra obtained in the experiments, and then used the AIC tests by calculating ($w_{EIT}, w_{ATS}$) and ($\bar{w}_{EIT}, \bar{w}_{ATS}$) proposed by Anisimov, Dowling, and Sanders[15] to determine which of the models (EIT or ATS) is the mostly likely for the experimental observation.

In **Fig. 5** we present typical curves of ($w_{EIT}, w_{ATS}$) and ($\bar{w}_{EIT}, \bar{w}_{ATS}$) obtained at three different experimental settings, corresponding to three different regimes of operation, as the coupling strength was increased: EIT-dominated regime (**Fig. 5a**), ATS-dominated (**Fig. 5b**) and EIT-to-ATS transition (**Fig. 5c**). The models assigned using AIC to the experimental data agree very well with the requirements to observe EIT or ATS.

In the first case (**Fig. 5a**), the WGMs in the resonators were chosen such that their decay rates, quantified by $Q_{\mu R1}$ (i.e., $\gamma_1 + \gamma_c$) and $Q_{\mu R2}$ (i.e., $\gamma_2$), significantly differed from each other (i.e., $Q_{\mu R2}/Q_{\mu R1} \sim 400$). We calculated $\kappa_T = |\gamma_1 + \gamma_c - \gamma_2|/4 = 312.8\,\text{MHz}$ which was larger than the coupling strengths used in the experiments ($\kappa < 100\,\text{MHz}$). Starting from $\bar{w}_{EIT} = \bar{w}_{ATS} = 0.5$ (i.e., both the EIT and the ATS models are equally likely) for very weak coupling strength $(0 \leq \kappa \leq 15\,\text{MHz} \ll \kappa_T)$, the likelihood of EIT model increased as the coupling strength was increased up to 100 MHz. Thus, we conclude that in this setting, which corresponded to weak driving regime ($\kappa < \kappa_T$), the data obtained in the experiments favors the EIT model, as revealed by $\bar{w}_{EIT} > \bar{w}_{ATS}$.

In the second case (**Fig. 5b**), the decay rates of the coupled WGMs were very similar to each other (i.e., $Q_{\mu R2}/Q_{\mu R1} \sim 0.5$), and we estimated the critical coupling strength as $\kappa_T = 16.2\,\text{MHz}$ which was smaller than the coupling strengths considered $\kappa > 60\,\text{MHz}$. Therefore, as demonstrated in the model, this falls in the strong-driving regime ($\kappa \gg \kappa_T$), where ATS is expected. Indeed, in this experimental setting, starting from $\bar{w}_{EIT} = \bar{w}_{ATS} = 0.5$, the likelihood of the ATS model increased as the coupling strength was increased up to 400 MHz. Thus, the data obtained in the experiments favors the ATS model as revealed by $\bar{w}_{ATS} > \bar{w}_{EIT}$ .

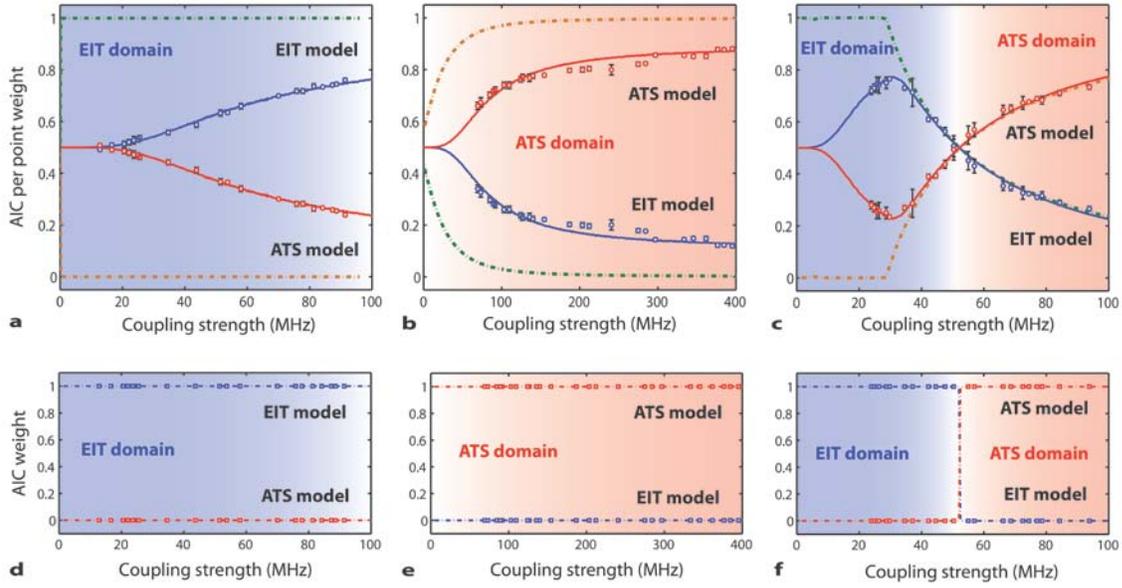

*Fig.5. Akaike-Information-Criterion (AIC) weights and per-point weights obtained as a function of the coupling strength in the coupled microcavities. (a) The AIC per-point weight for the pair of modes chosen in the first and second microresonators with Q ~ ($1.91 \times 10^5$, $7.26 \times 10^7$). As the coupling strength was increased, the EIT model dominates. (b) The AIC per-point weight for pair of modes with Q ~ ($1.63 \times 10^6$, $1.54 \times 10^6$). As the coupling strength was increased above 60 MHz, the ATS model dominates. (c) The AIC per-point weight for the pair of modes with Q ~ ($1.78 \times 10^6$, $4.67 \times 10^6$). A transition from EIT-dominated to ATS-dominated regime is clearly seen. (d)-(f) The AIC weights for the mode pairs studied in (a)-(c). The data for EIT and ATS are given in blue and red marks/curves, respectively, and the green and orange marks/curves represent the theoretical AIC per-point weights (i.e., these do not take into account the experimental noise).*

The third case (**Fig. 5c**) revealed an interesting phenomena: transition from an EIT-dominated regime to an ATS-dominated regime through an inconclusive regime, where both EIT and ATS are equally likely. The decay rates of the chosen WGMs were similar (i.e., $Q_{\mu R2}/Q_{\mu R1} \sim 2.6$); larger than that of the setting of **Fig. 5b,** but much smaller than that of the setting in Fig.5a. We estimated the critical coupling strength as $\kappa_T = 29.5\,\text{MHz}$. In this case, the EIT model first dominated ($\bar{w}_{EIT} > \bar{w}_{ATS}$) when the coupling strength was small. Then the likelihood of the EIT model decreased and that of the ATS model increased as the coupling strength was increased up to 50 MHz, where we observed $\bar{w}_{EIT} = \bar{w}_{ATS} = 0.5$. Further increase of the coupling strength beyond this point revealed a transition to an ATS-dominated regime ($\bar{w}_{ATS} > \bar{w}_{EIT}$). This results agree well with the predictions of the model: In the EIT-dominated regime we had $\kappa < \kappa_T$, in the transition regime we had $\kappa \sim \kappa_T$ and finally in the ATS-dominated regime we had $\kappa \gg \kappa_T$. In **Fig.5d-5f**, we also present ($w_{EIT}, w_{ATS}$) as a function of the coupling strength. As expected, these weights exhibit a binary behavior with an abrupt change from the EIT-dominated regime to the ATS-dominated regime.

Since we collected ten sets of data at each specific condition, we could assign standard deviations to the AIC weights as seen in **Fig. 5**. The technical noise in the experimental data points plays an accumulated role in the model fitting, which blurs the distinction between the models. This is clearly seen in the comparison of the AIC weights obtained from the experimental data with the theoretical weights. When the coupling strength was very small, in particular for the EIT regime, the noise had a larger blurring effect. This is attributed to the fact that in the very weak coupling regime, the EIT transparency window is so small that it is buried in the noise; thus the contribution of the transparency band to the whole fitting decreases. The factors that affect the fitting and hence the model-assignment according to AIC weights are thermal noise, the probe laser frequency and amplitude fluctuations, the limited number of data points acquired for each spectrum, the resolution of the oscilloscope and the efficiency of the detector. We estimated that the standard deviation of the total noise in our experiments is 1% to 2.5% of the measured signal.

Finally in **Fig.6** we show examples of typical transmission spectra obtained in our experiments in the EIT-dominated (**Fig.5a** and **6a**), the ATS-dominated (**Fig.5b** and **6b**) and the EIT-to-ATS transition regime (Fig.5c and 6c), together with the best-fitting curves using the expressions $T_{EIT}$ and $T_{ATS}$ derived from the theoretical model. It is clear that for the spectra for which the AIC assigned the EIT model, the function $T_{EIT}$ provided a better fit than $T_{ATS}$. In particular, the $T_{ATS}$ fitting performed poorly around the narrow transparency window (**Fig.6a** inset). In the spectrum for which the ATS model was assigned according to AIC weights, $T_{ATS}$ performed extremely well, whereas the $T_{EIT}$ fitting was very poor (**Fig.6b**). The experimental conditions for the data shown in **Fig.5c** revealed a transition from EIT to ATS. We chose a spectrum obtained in the vicinity of the transition point and performed curve fitting using $T_{EIT}$ and $T_{ATS}$. It is clearly seen in **Fig.6c** that $T_{EIT}$ and $T_{ATS}$ functions perform equally well and one cannot conclusively assign a model to it: We cannot conclusively show EIT (or ATS) nor rule EIT (or ATS) out. These curve-fitting tests (**Fig.6**) agree well with the predictions of the AIC weights (**Fig.5**).

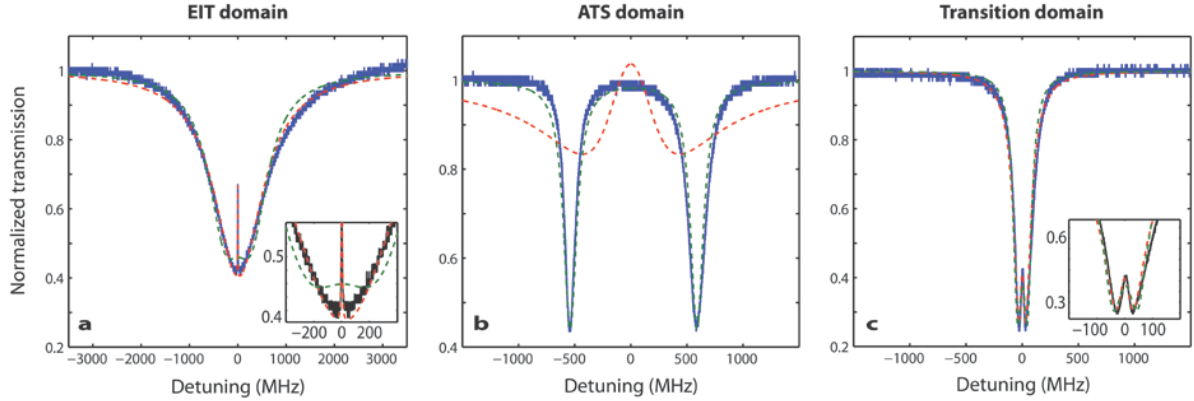

*Fig.6. Experimentally-observed transmission spectra with EIT and ATS model fittings. The transmission spectra shown here are chosen to represent the three regimes (EIT-dominated, ATS-dominated, and EIT-to-ATS transition regimes) observed in Fig.5 according to AIC weights and per-point weights. Data obtained in the experiments are given in blue curves (black in the insets), and the best fits using the $T_{EIT}$ model and the $T_{ATS}$ model are given by the red and green curves, respectively. (**a**) The mode pair with $Q \sim (1.91 \times 10^5, 7.26 \times 10^7)$ shows a clear transparency window, and the EIT model provides the best fit. This agrees with the result in Fig. 5a and 5d. (**b**) The mode pair with $Q \sim (1.78 \times 10^6, 4.67 \times 10^6)$ exhibits a broader transparency window with the best fit provided by the ATS model. This agrees with the result in Fig.5b and 5e. (**c**) At the transition region, both the EIT and the ATS models fit equally well to the transmission spectrum obtained for the mode pair with $Q \sim (1.63 \times 10^6, 1.54 \times 10^6)$. This agrees well with the results of Figs. 5c and 5f. Insets are enlarged versions of the spectra around the central frequency (zero-detuning).*

**Discussion**

Although initially proposed, observed and used in atomic and molecular systems, Fano interference, EIT and ATS are among many quantum phenomena which have classical and more importantly all-optical analogues. Their demonstrations in on-chip physical systems using optical microresonators, metamaterials or plasmonics offer great promises for a wide range of applications including controlling the flow of light on-chip, high performance sensors and studying the effects of many parameters which are difficult to test in atomic and molecular systems that need highly sophisticated and hard-to-access experimental environment and techniques. In particular, the capability of creating EIT and controlling the features of the transparency window in on-chip coupled optical microcavities is important for on-chip all-optical slowing and stopping of light, tunable optical filters, switching and nonlinear optics.

Our approach, demonstrated in this work, provides a highly accessible and controllable platform which allows to test the effects of all relevant parameters (coupling strength, decay rates, frequency detunings) on the same coupled-resonators system for the observation and control of Fano resonances, EIT and ATS, as well as to probe the transitions among them. The capability to fine-tune the parameters at a high level, as demonstrated here, enabled us to show the avoided-crossing in the ATS process as the frequency of one of the resonances were steered.

In order to make good use of the observed transparency windows in coupled microresonators for the practical applications mentioned above, it is crucial that we know whether Fano interferences have played a role or not (i.e., is the transparency the result of EIT or ATS?). Here we, for the first time, applied an objective test to characterize the parameters involved in Fano, EIT or ATS processes in coupled optical resonators and clearly discerned between EIT and ATS. The test used here is the Akaike information criterion proposed by Anisimov, Dowling, and Sanders[15]. This test clearly and with high confidence revealed whether the EIT or the ATS was involved in the experimentally-obtained transmission spectra under different operating conditions. In addition to its capability to discriminate between EIT and ATS, the test revealed the sensitivity of the parameters involved. Our study demonstrates the suitability of the AIC method to characterize EIT and ATS in coupled microresonator systems and to study the effects of the system parameters in the observed spectra in the transition regime.

**Acknowledgement**

This work was supported by the NSF under grant number 0954941 and the US Army Research Office under grant number W911NF-12-1-0026.

**APPENDIX A**

Here we give the results of numerical simulations depicting the expressions derived in the subsection *"Analogy between coupled optical resonators and three-level atoms"* of the main text for different driving regimes. As shown in the main text, for the coupled resonator system, the output field is given as $A_{out} = A_p + \sqrt{\gamma_c} A_1$ where the intracavity field $A_1$ can be written as $A_1 = i\sqrt{\gamma_c} A_p \chi$ with

$$\chi = \frac{(\omega + i\alpha_2)}{\kappa^2 - (\omega + i\alpha_1)(\omega + i\alpha_2)} \quad (3)$$

where we used $\alpha_k = i\omega_k + \gamma_k/2$ with $k = 1, 2$. This solution $\chi$ has a form similar to the response of an EIT medium (three-level atom) to a probe field. The normalized transmission $T = |A_{out}/A_p|^2$ is written as

$$T = 1 + \gamma_c^2 |\chi|^2 - 2\gamma_c \chi_i. \quad (4)$$

with $\chi_i$ representing the imaginary part of $\chi$. Since $1 \gg \gamma_c^2 |\chi|^2$ and $\gamma_c^2 |\chi|^2 \ll \gamma_c \chi_i$ we can re-write the transmission as

$$T = 1 - 2\gamma_c \chi_i \quad (5)$$

Clearly, it will be sufficient to analyze the behavior of $\chi_i$ only to understand the conditions leading to EIT or ATS. This is similar to considering the imaginary part of the susceptibility which determines the absorption of a probe in an atomic system.

We can re-write the expression in Eq.(3) as

$$\chi = \frac{\chi_+}{\omega - \omega_+} + \frac{\chi_-}{\omega - \omega_-} \quad (6)$$

where $\omega_\mp$ are the eigenfrequencies of the coupled system, $\chi_\pm = \mp(\omega_\pm + i\alpha_2)/\beta = -1/2 \pm i\xi/\beta$ satisfying $\chi_+ + \chi_- = -1$ and $\xi = (\gamma_1 - \gamma_2)/4$.

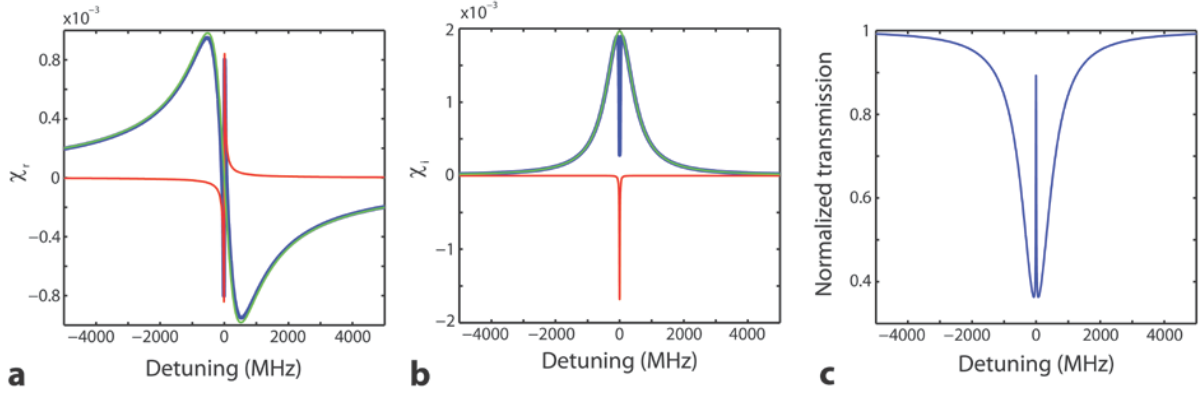

*Fig.A1 Real and imaginary parts of the susceptibility $\chi$ at the weak driving regime. (a) Real part of the susceptibility. (Blue: $\chi_{r1} + \chi_{r2}$, red: $\chi_{r1}$, green: $\chi_{r2}$). (b) Imaginary part of the susceptibility. (Blue: $\chi_{i1} + \chi_{i2}$, red: $\chi_{i1}$, green: $\chi_{i2}$). (c) Normalized transmission. The parameters used in (a)-(c) were obtained in the experiments: Decay rate of the first resonator was $\gamma_1 = 1.05 GHz$; decay rate of the second resonator was $\gamma_2 = 3 MHz$ and coupling strength was $\kappa = 67 MHz$.*

*(a) Weak-driving regime ($\kappa < \kappa_T$).* In this regime $\beta$ is imaginary, that is $\beta = i\beta_i$ and $\mathrm{Re}(\beta) = \beta_r = 0$ leading to real $\chi_\pm$ (i.e., $\mathrm{Im}(\chi_\pm) = \chi_{\pm i} = 0$) with $\mathrm{Re}(\chi_\pm) = \chi_{\pm r} = -1/2 \pm \xi/|\beta|$, and imaginary eigenfrequencies (i.e., $\mathrm{Re}(\omega_\pm) = \omega_{\pm r} = 0$) with $\mathrm{Im}(\omega_\pm) = \omega_{\pm i} = -\zeta \pm |\beta|$ where $\zeta = (\gamma_1 + \gamma_2)/4$. Thus the supermodes have the same resonance frequencies and are located at the center of the frequency axis, but they have different linewidths quantified by their imaginary parts. As a result, the real and imaginary parts of $\chi$ become

$$\chi_r = \frac{\omega \chi_{+r}}{\omega^2 + \omega_{+i}^2} + \frac{\omega \chi_{-r}}{\omega^2 + \omega_{-i}^2} \tag{7}$$

$$\chi_i = \frac{\omega_{+i} \chi_{+r}}{\omega^2 + \omega_{+i}^2} + \frac{\omega_{-i} \chi_{-r}}{\omega^2 + \omega_{-i}^2} \tag{8}$$

from which we write the transmission as

$$T_{EIT} = 1 - 2\gamma_c \chi_i = 1 - 2\gamma_c \left[ \frac{\omega_{+i} \chi_{+r}}{\omega^2 + \omega_{+i}^2} + \frac{\omega_{-i} \chi_{-r}}{\omega^2 + \omega_{-i}^2} \right] = 1 - \frac{C_1}{\omega^2 + \Gamma_1^2} + \frac{C_2}{\omega^2 + \Gamma_2^2}. \quad (9)$$

We have plotted the expressions in Eqs. (7)-(9) in Fig. A1 which shows that the normalized transmission becomes maximum around the zero frequency-detuning exactly where the imaginary part of $\chi$ becomes minimum. Clearly the transmission behavior of the system is determined by the $\chi_i$.

**(b) Strong-driving regime ($\kappa \gg \kappa_T$).** In this regime $\beta = 2\kappa$ is real (i.e., $\beta_i = 0$ and $\beta_r = 2\kappa$) implying $\omega_\pm = -i\zeta \pm \kappa$, that is the resonances are located at frequencies $\pm \kappa$ with a spectral distance of $2\kappa$ between them. The linewidths of the resonances are quantified by $\text{Im}(\omega \pm) = -\zeta/4$. Moreover, we can approximate $\chi_\pm$ as $\chi_\pm = -1/2$. Then we can write the imaginary parts of $\chi$ as

$$\chi_r = -\frac{1}{2}\left[ \frac{\omega - \kappa}{(\omega - \kappa)^2 + \zeta^2} + \frac{\omega + \kappa}{(\omega + \kappa)^2 + \zeta^2} \right] \quad (10)$$

$$\chi_i = \frac{1}{2}\left[ \frac{\zeta}{(\omega - \kappa)^2 + \zeta^2} + \frac{\zeta}{(\omega + \kappa)^2 + \zeta^2} \right] \quad (11)$$

Consequently, the transmission in this regime becomes

$$T_{ATS} = 1 - \gamma_c \left[ \frac{\zeta}{(\omega - \kappa)^2 + \zeta^2} + \frac{\zeta}{(\omega + \kappa)^2 + \zeta^2} \right] = 1 - \frac{C}{(\omega - \delta_0)^2 + \Gamma^2} - \frac{C}{(\omega - \delta_0)^2 + \Gamma^2} \quad (12)$$

with $\delta_0 \approx \pm \kappa$. We have plotted $T_{ATS}$ using the parameters obtained in the experiments and depicted it in Fig. A2. Clearly, imaginary part of the susceptibility function determines the normalized transmission which consists of two Lorentzian resonances separated by $2\kappa$. We should note that $\chi_i$ shows Lorentzian peaks whereas the normalized transmission shows Lorentzian dips with a larger transparency window than that of the transmission spectra obtained for the weak-driving regime (EIT case).

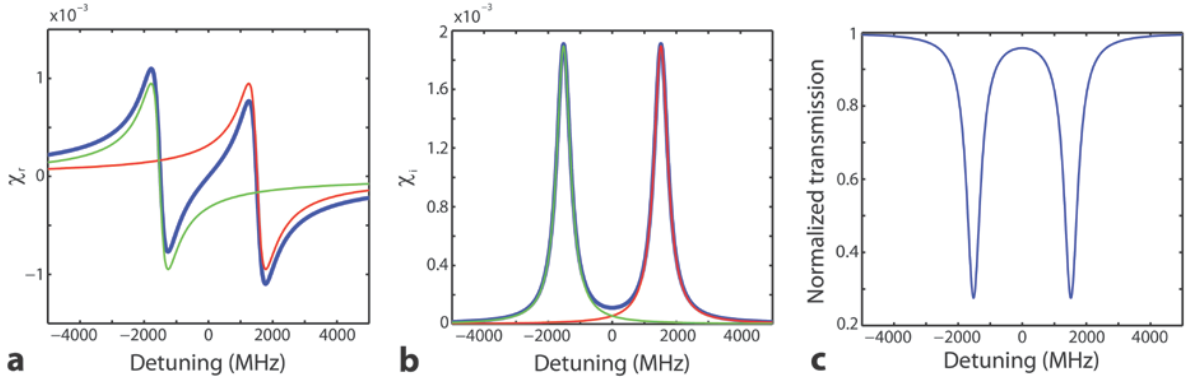

***Fig.A2 Real and imaginary parts of the susceptibility*** $\chi$ ***at strong driving regime.*** (***a***) *Real part of the susceptibility.* (*Blue:* $\chi_{r1} + \chi_{r2}$, *red:* $\chi_{r1}$, *green:* $\chi_{r2}$). (***b***) *Imaginary part of the susceptibility.* (*Blue:* $\chi_{i1} + \chi_{i2}$, *red:* $\chi_{i1}$, *green:* $\chi_{i2}$). (***c***) *Normalized transmission. The parameters used in (a)-(c) were obtained from experiments: Decay rate of the first resonator was* $\gamma_1 = 1.05 GHz$; *decay rate of the second resonator was* $\gamma_2 = 3 MHz$; *coupling strength was* $\kappa = 1.52 GHz$

*(c) Intermediate-driving regime* ($\kappa > \kappa_T$). In this regime $\beta = \beta_r$ is real (i.e., $\beta_i = 0$). This leads to complex eigenfrequencies $\omega_\pm = (-i\gamma_1 - i\gamma_2 \pm 2\beta_r)/4$ (i.e., $\text{Re}(\omega_\pm) = \omega_{\pm r} \neq 0$ and complex $\chi_\pm$ (i.e., $\text{Im}(\chi_\pm) = \chi_{\pm i} \neq 0$). Then $\text{Im}(\chi_\pm) = \chi_{\pm i} = \pm(\gamma_1 - \gamma_2)/4|\beta|$ with $\text{Re}(\chi_\pm) = \chi_{\pm r} = -1/2$. The real and imaginary parts of the eigenfrequencies are $\text{Re}(\omega_\pm) = \omega_{\pm r} = \pm\beta/2$ and $\text{Im}(\omega_\pm) = \omega_{\pm i} = (-\gamma_1 - \gamma_2)/4$, respectively. Thus the supermodes have different resonance frequencies which are located at $\pm\beta/2$, but have the same linewidths quantified by their imaginary parts $\text{Im}(\omega_\pm) = \omega_{\pm i} = (-\gamma_1 - \gamma_2)/4$. Consequently, we the real and imaginary parts of $\chi$ become

$$\chi_r = \frac{(\omega - \omega_{+r})\chi_{+r} - \omega_{+i}\chi_{+i}}{(\omega - \omega_{+r})^2 + \omega_{+i}^2} + \frac{(\omega - \omega_{-r})\chi_{-r} - \omega_{-i}\chi_{-i}}{(\omega - \omega_{-r})^2 + \omega_{-i}^2} \qquad (13)$$

$$\chi_i = \frac{(\omega - \omega_{+r})\chi_{+i} + \omega_{+i}\chi_{+r}}{(\omega - \omega_{+r})^2 + \omega_{+i}^2} + \frac{(\omega - \omega_{-r})\chi_{-i} + \omega_{-i}\chi_{-r}}{(\omega - \omega_{-r})^2 + \omega_{-i}^2} \qquad (14)$$

which imply that $\chi_i$ is the sum of two same-sign quasi-Lorentzians centered at $\pm\beta/2$. The transmission at this region is then given as

$$T_{EIT/ATS} = 1 - 2\gamma_c \chi_i = 1 - 2\gamma_c \left[ \frac{(\omega - \omega_{+r})\chi_{+i} + \omega_{+i}\chi_{+r}}{(\omega - \omega_{+r})^2 + \omega_{+i}^2} + \frac{(\omega - \omega_{-r})\chi_{-i} + \omega_{-i}\chi_{-r}}{(\omega - \omega_{-r})^2 + \omega_{-i}^2} \right]$$

$$= 1 - 2\gamma_c \left\{ \left[ \frac{(\omega - \omega_{+r})\chi_{+i}}{(\omega - \omega_{+r})^2 + \omega_{+i}^2} + \frac{(\omega - \omega_{-r})\chi_{-i}}{(\omega - \omega_{-r})^2 + \omega_{-i}^2} \right] + \left[ \frac{\omega_{+i}\chi_{+r}}{(\omega - \omega_{+r})^2 + \omega_{+i}^2} + \frac{\omega_{-i}\chi_{-r}}{(\omega - \omega_{-r})^2 + \omega_{-i}^2} \right] \right\} \quad (15)$$

$$= 1 - \left\{ \left[ \frac{(\omega - \varepsilon)C_1}{(\omega - \varepsilon)^2 + \Gamma^2} - \frac{(\omega + \varepsilon)C_1}{(\omega + \varepsilon)^2 + \Gamma^2} \right] + \left[ \frac{C_2}{(\omega - \varepsilon)^2 + \Gamma^2} + \frac{C_2}{(\omega + \varepsilon)^2 + \Gamma^2} \right] \right\}$$

where the second expression in the bracket is similar to the expression derived for the strong-driving regime (ATS case). Thus we re-write Eq. (15) as

$$T_{EIT/ATS} = T_{ATS} - C_1 \left[ \frac{(\omega - \varepsilon)}{(\omega - \varepsilon)^2 + \Gamma^2} - \frac{(\omega + \varepsilon)}{(\omega + \varepsilon)^2 + \Gamma^2} \right] \quad (16)$$

Figure A3 depicts the transmission function given in Eq. 16 together with the real and imaginary parts of the susceptibility.

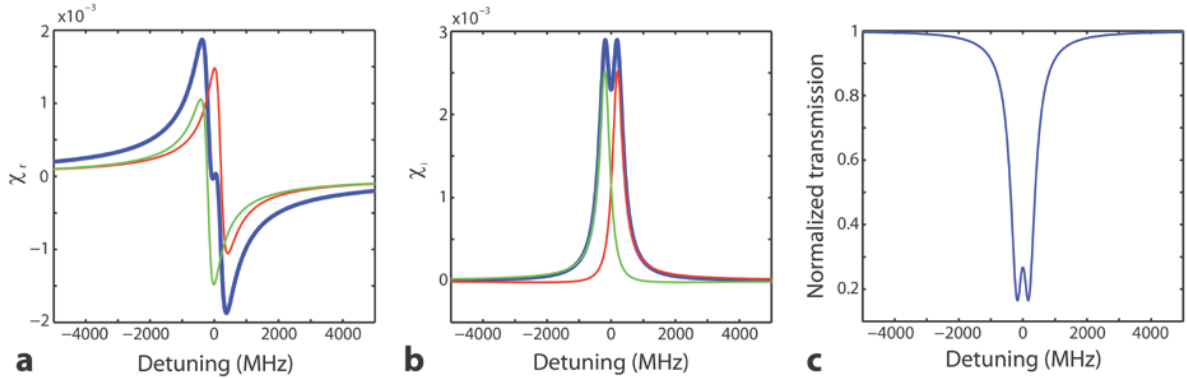

*Fig.A3 Real and imaginary parts of the susceptibility $\chi$ at intermediate-driving regime. (a) Real part of the susceptibility. (Blue: $\chi_{r1} + \chi_{r2}$, red: $\chi_{r1}$, green: $\chi_{r2}$). (b) Imaginary part of the susceptibility. (Blue: $\chi_{i1} + \chi_{i2}$, red: $\chi_{i1}$, green: $\chi_{i2}$). (c) Normalized transmission. The parameters used were obtained from experiments. Decay rate of the first resonator was $\gamma_1 = 462 MHz$; decay rate of the second resonator was $\gamma_2 = 337 MHz$; coupling strength was $\kappa = 186 MHz$.*

## APPENDIX B

In the main text, although we have derived the normalized transmission for weak, strong and intermediate driving regimes, in the model selection problem we used only the expressions for EIT (weak driving regime) and ATS (strong driving regime). The reason behind this was that the model for the intermediate-driving regime EIT/ATS contains two terms: One is exactly the same as the expression derived for the strong driving regime (ATS) and the other is an interference term whose contribution can be set to zero or minimized by properly choosing the coupled modes or is set to zero or much lower values than the contribution from the ATS part during the curve-fitting algorithm due to the fact that $C_1$ is a free-parameter. Here we give the results of our study in which we performed curve fitting using the EIT, ATS and EIT/ATS models to the calculated theoretical transmission spectra obtained using experimentally relevant parameter values. In the transmission spectra we also included 1% Gaussian noise. Moreover, we give the AIC per-point weights for the three driving regimes.

We have observed that the ATS and the EIT/ATS models have the same AIC per-point weights. Results obtained for typical transmission spectra are depicted in Fig. B1. As the coupling strength increases, the $w_{ATS}$ and $w_{EIT/ATS}$ exhibit the same values. This is expected as we have mentioned above that the transmission in the intermediate-driving regime includes the contribution from the ATS model as shown in Eq. (16). Therefore, the $w_{ATS}$ and $w_{EIT/ATS}$ are always have similar values as the system evolves from weak to strong driving regimes.

We also checked the difference between the ATS (strong driving regime) and the EIT/ATS (intermediate driving regime) models by calculating the AIC weights without the averaging effect. The binary weights in Fig. B2 reveals that the AIC weights of the EIT/ATS model is always lower than those of ATS model, because although the EIT/ATS model gives a more precise fitting to the experimental data, the higher number of free parameters in the EIT/ATS model adds larger cost (i.e., increased penalty) and hence reduces the weights of this model.

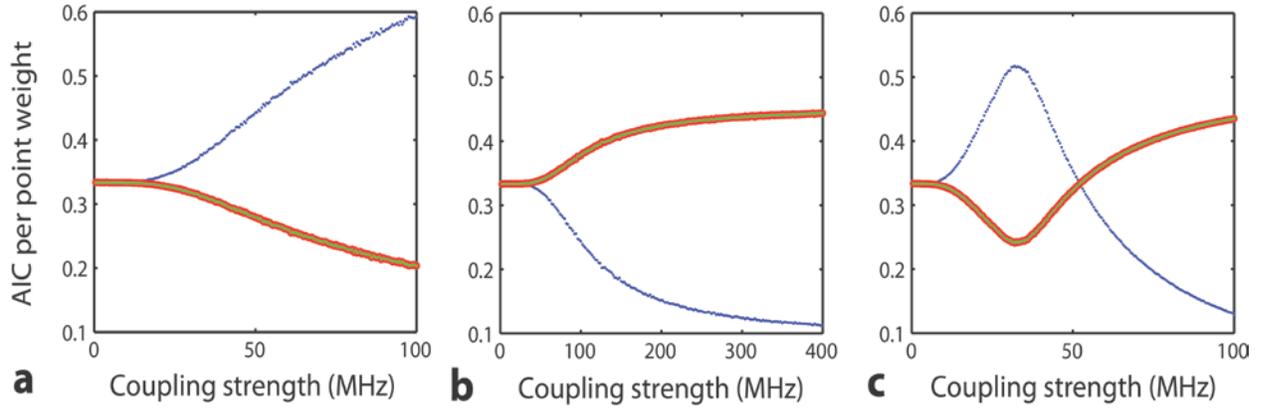

*Fig.B1 Theoretical (noise model) AIC per-point weights as the function of coupling strength for EIT, ATS, and intermediate-driving models. (Blue: $w_{EIT}$, green: $w_{EIT/ATS}$, red and thicker: $w_{ATS}$) (**a**) The AIC per-point weight for the mode pair with Q~$(1.91\times10^5, 7.26\times10^7)$. (**b**)The AIC per-point weight for the mode pair with Q~$(1.63\times10^6, 1.54\times10^6)$. (**c**) The AIC per-point weight for the mode pair with Q~$(1.78\times10^6, 4.67\times10^6)$.*

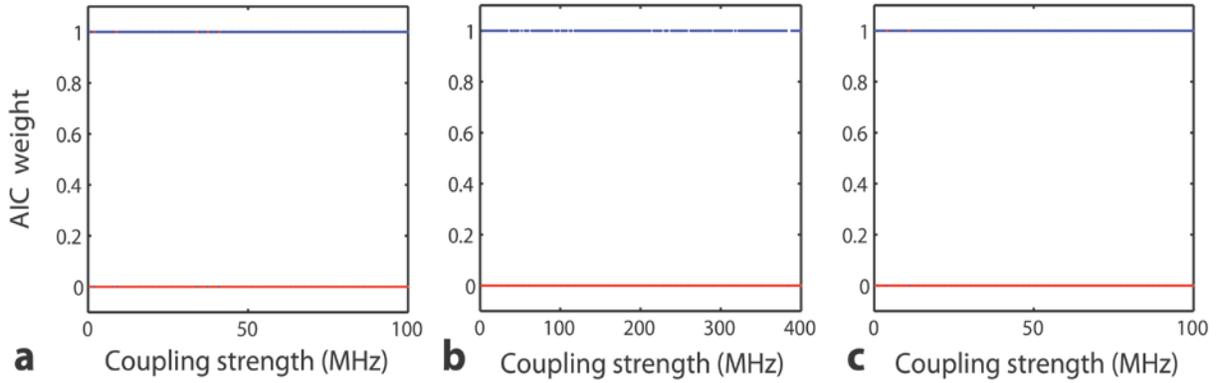

*Fig.B2 Theoretical (noise model) AIC weights as the function of coupling strength for ATS and EIT/ATS models. (Blue: $w_{ATS}$, red: $w_{EIT/ATS}$) (**a**) The AIC per-point weight for the mode pair with Q~$(1.91\times10^5, 7.26\times10^7)$. (**b**) The AIC per-point weight for the mode pair with Q~$(1.63\times10^6, 1.54\times10^6)$. (**c**) The AIC per-point weight for the mode pair with Q~$(1.78\times10^6, 4.67\times10^6)$.*

Finally, we used the intermediate driving model (EIT/ATS) to fit to the typical transmission spectra obtained in our experiments. The results are depicted in Fig. B3. It is seen that for the transmission spectra for EIT case (same as the one in the main text), the EIT/ATS model do not provide a good fit although it has more free parameters. The discrepancy is significant around the zero-detuning where we have the transparency window. For the spectra obtained for the ATS and EIT-to-ATS transition, we see that the EIT/ATS model provides a curve-fitting at least as good as the ATS model.

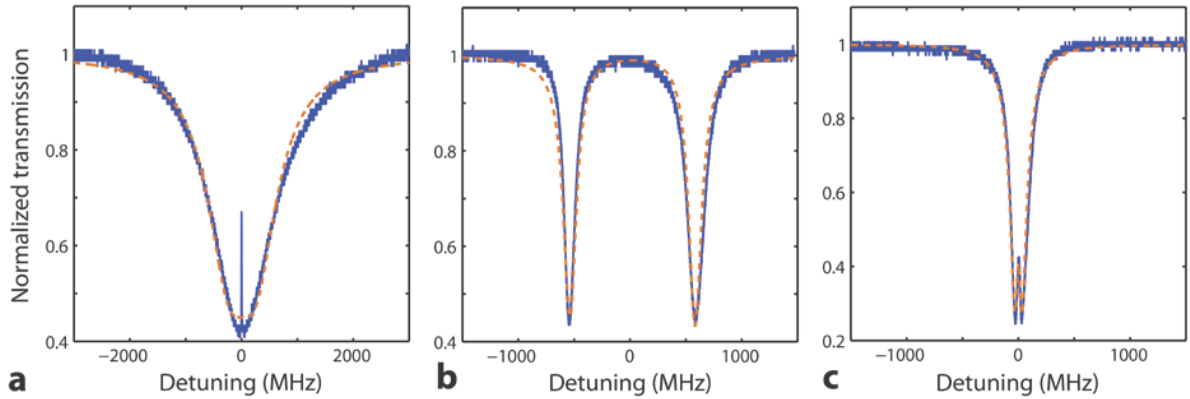

*Fig.B3 Fitting with the intermediate-driving model to the experimentally obtained spectra. Blue: experiment, Orange: Intermediate-driving model fitting. (**a**) The fitting spectra for the mode pair with Q~($1.91\times10^5, 7.26\times10^7$). (**b**) The AIC fitting spectra for the mode pair with Q~($1.63\times10^6, 1.54\times10^6$). (**c**) The fitting spectra for the m*